# Mechanisms governing resonant shifts and resilience in optical responses of plasmonic hole arrays

Amir Djalalian-Assl

*51 Golf View Drive, Craigieburn, VIC 3064*

This report concerns with the theoretical studies of SPP Bloch waves in arrays of cylindrical holes. Most analytical solutions are simplified, approximate and fitted, thus leading to wrong design parameters. A rigorous analytical solution that can accurately predict the desired dimensions associated with hole arrays is complex to derive. Finite element analysis (FEA), on the other hand, offers extensive capabilities for modelling and simulating the response of periodic structures such as that of an infinite array of holes. In this chapter, a square array of cylindrical holes is modelled numerically. A simple analytical model is presented (not for design purposes) only to identify the origin of the *alleged* (1,1) and (0,2) modes and the results are compared to the numerical ones. Changes in the transmission spectrum vs the refractive index of the surrounding dielectric when illuminated with a linearly polarized light at normal incidence is studied. Circumstances in which an obliquely incident light changes to the SPP momentum are also identified.

**Introduction**

Various attempts have been made to explain the resonance shift and the Fano profile associated with the extraordinary optical transmission (EOT) through subwavelength hole arrays in terms of the coupling of SPP Bloch waves to the non-resonant scattering process from the holes [1-4]. The role of resonant and non-resonant Wood anomalies in shaping the spectral line was also investigated [5]. Recent investigations on the origin of EOT in metallic hole arrays suggest that the contributions of SPPs are limited to only 50% of the total EOT, with quasi-cylindrical waves (QCW) being responsible for the other 50% [6-8]. The optical transmission associated with hole arrays in metallic films can be described in terms of resonant excitation of SPP modes responsible for enhanced transmissions and Wood's anomalies [9] that are responsible for suppressed transmissions. Understanding SPP-Bloch waves and Wood's anomalies in aperture antennas is important when designing antenna arrays which rely solely on the cavity mode. The SPP-LSP coupling may or may not be a desirable effect depending on the requirements. When designing plasmonic antenna arrays, however, one cannot rely on the analytical relations governing the surface modes as I will demonstrate in this report.

**Modelling, Results and Analysis**

An array of cylindrical holes was modelled using a 100 nm silver film set in the *x-y* plane, sandwiched between two semi-infinite dielectric slabs with refractive indices $n_1$ and $n_2$. The dielectric filling the holes has an index of $n_3$. The structure was normally illuminated with a plane wave propagating in the $+z$ direction, polarized at 0° to the *x* axis at $\lambda_0 = 700$ nm. To prevent transmission through the film and propagating modes through the holes at $\lambda_0 = 700$ nm, the thickness of the silver film was set to $h = 100$ nm and the diameter of the holes to $d = 200$ nm [10]. Note that the skin depth for silver is $\delta \approx 25$ nm in the visible regime. The relative permittivity of bulk silver at the design wavelength of $\lambda_0 = 700$ nm was calculated from tabulated experimental data to be $\varepsilon_{Ag}(\lambda = 700nm) = \varepsilon'_{Ag} - i\varepsilon''_{Ag} = -20.43279 - 1.2862i$ [11]. Various techniques, such as perfectly matched layers, scattering boundary conditions and ports were employed to eliminate the back reflection of the diffracted wave from the upper and the lower boundaries of the model. Terminating the upper and the lower boundaries with ports was found to be the most appropriate approach for square arrays. In this section, the silver film was assumed to be supported by a glass with a refractive index of 1.52. The simulated absolute transmission (normalized to the incident intensity over a unit cell) as a function of the periodicity at $\lambda_0 = 700$ nm for $n_1 = 1.52$ and $n_2 = n_3 = 1$, (i.e. silver hole array on a glass substrate), is shown in Figure 1(a). A maximum is observed at the periodicity of $P = 394$ nm. The absolute transmission vs. the wavelength for numerically obtained periodicity $P = 394$ nm is depicted in Figure 1(b). The numerical solution produced the modes at $(1,0)_{\text{W-glass}} = 650$ nm, $(1,1)_{\text{W-glass}} = 500$ nm and $(1,0)_{\text{SPP-glass}} = 700$ nm. Although the peaks at 540 nm and 475 nm are being *attributed* to the $(1,1)_{\text{SPP-glass}}$ and $(2,0)_{\text{SPP-glass}}$ respectively, it can be shown that this is not the case.

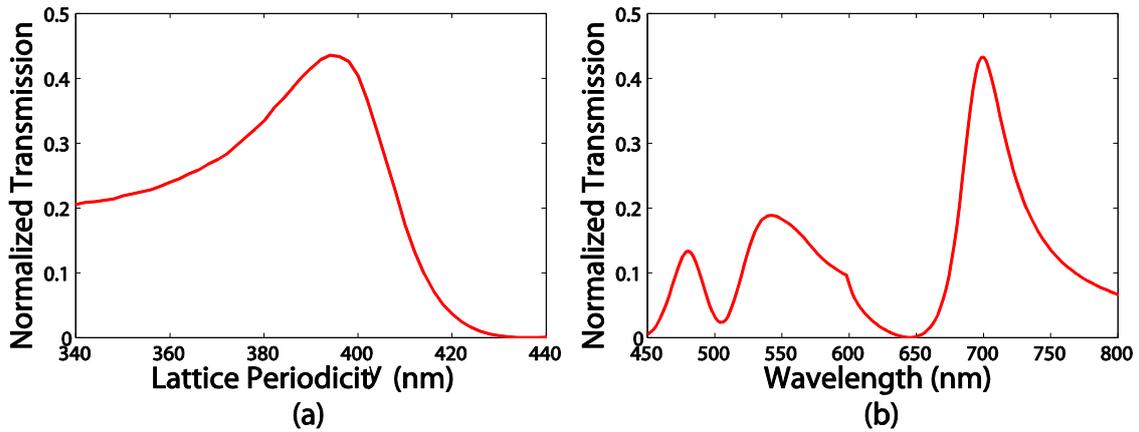

**Figure 1:** Absolute transmission spectra (normalized to the incident intensity over a unit cell) through a square array of holes as a function of (a) the periodicity at $\lambda_0 = 700$ nm and (b) the wavelength for $P = 394$ nm,

In a square array the condition for Wood's anomalies are given by [9]:

$$P = \frac{\lambda_W}{\sqrt{\varepsilon_d}}\sqrt{i^2 + j^2} \qquad (1)$$

And SPP Bloch waves on the surface of a metallic hole array follow the conservation of momentum [12-14]:

$$\mathbf{k}_\parallel \pm i\mathbf{G}_x \pm j\mathbf{G}_y = \mathbf{k}_{SPP}$$

$$k_{SPP} = Re\left(\sqrt{\frac{\varepsilon_m \varepsilon_d}{\varepsilon_m + \varepsilon_d}}\frac{2\pi}{\lambda_0}\right) \qquad (2)$$

$$P = \frac{2\pi}{k_{SPP}}\sqrt{i^2 + j^2} \qquad (3)$$

where $\varepsilon_m$ and $\varepsilon_d$ are the complex permittivities of the metal and the dielectric respectively, and $i$ and $j$ are integers and $\lambda_W$ is the free space wavelength associated with the Wood's anomaly.

Using equations (2) and (3), for $i = 0$ and $j = 1$, the periodicity that supports the fundamental SPP Bloch mode at $\lambda_0 = 700$ nm is analytically calculated to be $P = 2\pi/G = 433$ nm. Using equations (1), and (3), analytical values for the free space wavelengths associated with the resonant SPP Bloch modes and Wood's anomalies that fall within the optical spectrum for $P = 433$ nm were calculated to be $(1,0)_{\text{SPP-glass}} = 700$ nm, $(1,1)_{\text{SPP-glass}} = 530$ nm, $(2,0)_{\text{SPP-glass}} = 433$ nm, $(1,0)_{\text{W-glass}} = 658$ nm and $(1,1)_{\text{W-glass}} = 465$ nm. Although the simulated $(1,1)_{\text{SPP-glass}}$ mode seems to agree with the corresponding analytical value, large difference observed between the simulated $(2,0)_{\text{SPP-glass}}$ mode and that obtained analytically, raises suspicions that other phenomena may be involved in shaping the spectral lines.

Although it seems that values obtained analytically are in agreements with the simulations, one cannot rely on the analytical solution for design purposes. Figure 2 depicts the absolute transmission spectra for the same array with various combinations of refractive indices of the substrate, superstrate and the hole. To segregate and identify modes supported at glass/silver interface from those supported at air/silver interface, the transmission spectra vs. the wavelength for $n_1 = n_2 = n_3 = 1$ was simulated. The spectrum shows that SPP Bloch modes at the air/silver interface have a cut-off wavelength of $\lambda_0 = 530$ nm, corresponding to the $(1,0)_{\text{SPP-air}}$, and is well away from the target wavelength $\lambda_0 = 700$ nm. The $n_1 = n_2 = 1.52$ spectrum, with the index of the material filling the holes kept fixed at $n_3 = 1$, shows a redshift in the fundamental mode from $\lambda_0 = 700$ nm to $\lambda_0 = 730$ nm caused by the presence of the glass at both the incident and the exit interfaces. The $(1,1)_{\text{SPP-glass}}$ mode, on the other hand, maintained its original position at 530 nm. This implies that an extra momentum could not have been responsible for shifting the $(1,0)_{\text{SPP-glass}}$ resonance. Changes in the amplitude and the position of the mode, in this case, may be explained in terms of the coupling of surface modes to the Fabry-Pérot (FP) standing waves inside the holes. Resonant SPP Bloch waves as well as the FB resonance of the holes are influenced by the film thickness, array periodicity and the diameter of the holes, hence the Fabry-Pérot Evanescent Wave (FPEV) [15]. The offset between the two resonances influences both the peak wavelength and the enhancement/suppression of EOT. With $n_1 = n_3 = 1.52$ and the superstrate being the air, the $(1,0)_{\text{SPP-glass}}$ is redshifted to $\lambda_0 = 760$ nm. For $n_1 = n_2 = n_3 = 1.52$, a further shift to $\lambda_0 = 800$ nm is observed. In both cases with $n_3 = 1.52$, the $(1,1)_{\text{SPP-glass}}$ mode was also redshifted away from its original peak wavelength of 530 nm.

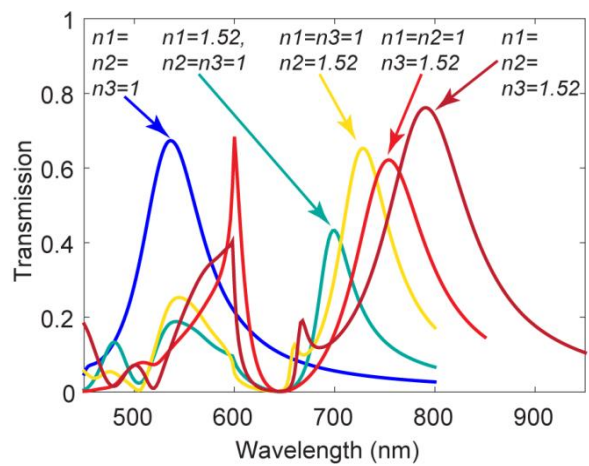

**Figure 2: Transmission (normalized to the incident intensity over a unit cell area) spectra of a square array of holes, 200 nm in diameter and periodicity of 394 nm, perforated in 100 nm silver film, for various combinations of refractive indices of the substrate, superstrate and the hole.**

Whether subwavelength cylindrical holes support guided modes, is a matter of dispute. Ebbesen and co-workers are inclined towards a cut-off wavelength of $\lambda_0 = d/2$ for guided modes through cylindrical holes [10] hence attributed the origin of the extraordinary optical transmission to the coupling of the light and the SPPs on the incident surface [13] that initiate an evanescent tunneling process through the holes. In a later publication [16], they changed their claim to include the exit surface as part of the tunneling process without any justification. And yet this was done in the same article showing that the coupling between the incident light and the SPP Bloch waves at the silver/air interface has a different EOT dispersion curve in comparison to that of the silver/glass interface [16]. Indeed, such claims have created confusion and misunderstandings among the plasmonic community that I intend to clarify here. I tend to agree with Catrysse *et. al.* that proved subwavelength cylindrical holes do support guided modes [17-19]. For a hole array perforated in an optically thick metallic film the dispersion relation at the metal/air interface as a function of the angle of incidence, is traced along a separate band structure in comparison to that of the metal/glass interface [20]. As previously mentioned: the cut-off wavelength of $\lambda_0 = 530$ nm, corresponding to the $(1,0)_{SPP\text{-}air}$ Bloch mode, is well away from the target wavelength $\lambda_0 = 700$ nm. This implies that if the array is illuminated by a normally incident light at $\lambda_0 = 700$ nm from the air/silver interface, none of the Bloch modes at the air/silver interface would be excited. Consequently, in the absence of SPPs on the incident surface, the so-called "evanescent tunneling process through the subwavelength holes" claimed by Ebbesen and co-workers, is never initiated. Therefore, one must not expect the formation of SPP Bloch waves on the exit (glass/silver) surface either. However, my simulations showed the formation of SPPs at the glass/silver interface with no (or faint) SPPs at the air/silver interface when the array was illuminated from the air with a normally incident light at $\lambda_0 = 700$ nm. This is only possible if the holes have transmitted the incident light without the assistance of the (frail or non-existent) SPPs at the incident air/silver interface, Figure 3. This may provide us with a solution for minimizing the depolarization effect in hole arrays which may result from the off-normal incidence.

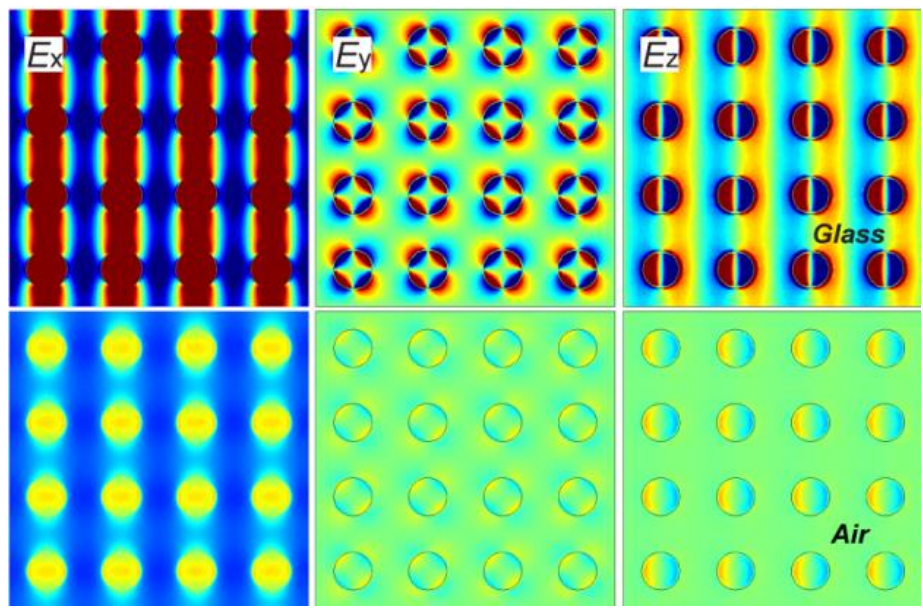

**Figure 3: Electric field components of the SPPs calculated over the silver/glass surface (top row) and air/silver interface (bottom row) over the array of holes 200 nm in diameter perforated in a 100 nm thick silver film. Period of the array is $P = 433$ nm. The array was illuminated with a normally incident *x*-polarized light from the air/silver interface. Note that $E_x$ for the air/silver interface (bottom-left) depicts the superposition of the incident field and the SPP field. All images were produced on the same scale with (red = 1, blue = -1 and green = 0).**

It is well known that in the case of an off-normal incidence on metallic surfaces, the parallel to the surface component of the incident wavevector makes a contribution to the resultant SPP wavevector, see equation (2), which impacts the state of polarization at transmission, hence the term *depolarization* [21, 22]. Reiterating the fact that the period of the above array was designed based on the fundamental surface mode supported at the glass/silver interface, it is obvious that the parallel-to-the-surface component of the incident wavevector contribute to the momentum of $(i,j)_{SPP-glass}$ modes only if the light is incident on the glass/silver interface. Figure 4 depicts the absolute transmission (normalized to the intensity over a unit cell) calculated in the air half-space vs. the angle of incidence, θ, when the array is illuminated from the glass half-space with a TE polarized light in the *x-z* plane, i.e. $\mathbf{E} \cdot \mathbf{k}_{\parallel} = 0$ for all θ. Any changes observed in the spectrum, therefore, are purely due to the contribution that $\mathbf{k}_{\parallel} = \mathbf{k}_0 \sin(\theta)$ makes to the SPP momentum. For θ = 10°, the resonance wavelength associated with the $(1,0)_{SPP-glass}$ mode, is blue shifted from $\lambda_0$ = 700 nm to $\lambda_0$ = 690 nm, whereas the analytical solution, see equation (2), predicts the resonance shift to $\lambda_0$ = 640 nm or $\lambda_0$ = 760 nm for $(+1,0)_{SPP-glass}$ and $(-1,0)_{SPP-glass}$ respectively. A +10 nm shift in resonance, instead of ±60 nm, cannot be explained by the equation (2). The coupling of the parallel-to-the-surface component of the incident "electric" field to SPP modes is influential in shaping the spectra of the hole array when excited with an oblique incident light. Figure 5 shows the calculated transmission through the array when it is illuminated from the substrate with a TM polarized light, i.e. $\mathbf{E} \cdot \mathbf{k}_{\parallel} > 0$, for θ > 0. The redshift in $(-1,0)_{SPP-glass}$ resonance and blue-shift in the degenerate mode $(+1,0)_{SPP-glass}$ are now in accordance with equation(2). The amplitude of the degenerate mode, however, builds up gradually with increasing θ. The only difference between the TE and TM mode, is the presence of the *z*-component of the incident electric field in the latter. So it seems that it is the *z*-component of the incident field that interacts with the hole, giving rise to $\mathbf{k}_{\parallel}$. Clearly equation (2) has no provision for TE vs. the TM illuminations since it only addresses the SPP Bloch waves. Please also see the Transmission spectra through a hole array vs. the incident angle reported by Ebbesen *et.al*[13] for comparison.

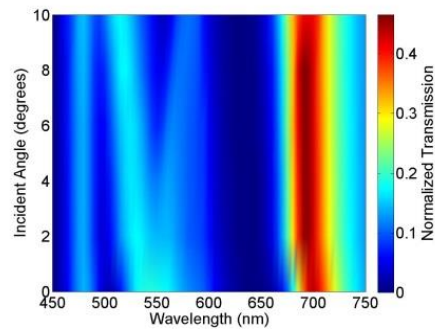

**Figure 4: Absolute transmission (normalized to the intensity over a unit cell) vs. the angle of incidence when the device is illuminated from the glass/silver side. TE mode, with *x-z* being the plane of incidence.**

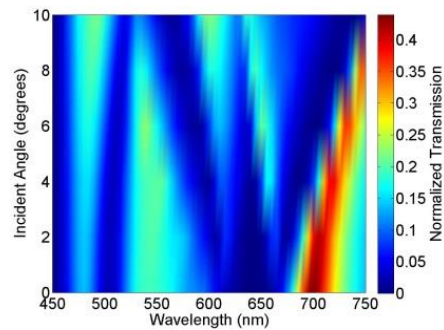

**Figure 5: Absolute transmission (normalized to the intensity over a unit cell) vs. the angle of incident when the device is illuminated from the glass/silver side. TM mode, with *x-z* being the plane of incidence.**

The other scenario where the off-normal light is incident on the air/silver interface, the (i,j)$_{SPP-glass}$ modes (that are excited only by the extraordinary transmitted light through the holes) are shielded from the incident light and desensitized to the angle of incidence. In such a scenario only the (i,j)$_{SPP-air}$ modes couple to the incident light and since the cut-off wavelength for (i,j)$_{SPP-air}$ is at $\lambda_0$ = 530 nm, the (1,0)$_{SPP-glass}$ mode at $\lambda_0$ = 700 nm remains uninfluenced, compare Figure 5 to Figure 7. In the case of the off-normal incidence in TE mode, **k**$_{SPP}$ and the parallel-to-the-surface component of the incident electric field are orthogonal, therefore there will be negligible changes to the SPP's momentum, see Figure 4 and Figure 6.

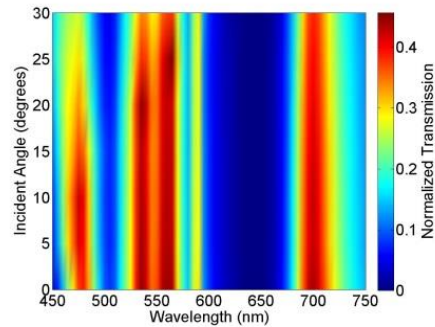

**Figure 6: Absolute transmission (normalized to the intensity over a unit cell) vs. the angle of incident when the device is illuminated from the air/silver side. TE mode, with *x-z* being the plane of incidence.**

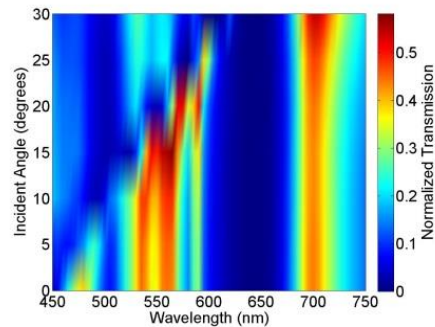

**Figure 7: Absolute transmission (normalized to the intensity over a unit cell) vs. the angle of incident when the device is illuminated from the air/silver side. TM mode with *x-z* being the plane of incidence.**

## Conclusion

The contribution of the actual (1,1)$_{SPP-glass}$ mode to the overall transmission spectrum of a square array of holes was found to be insignificant. According to the analytical mode, the origin of the conventionally labelled (1,1), (0,2) modes were attributed to the constructive interference of the complex superposed standing waves formed between the nearest neighboring holes that did not include the diagonal modes, as I have elaborated in arXiv:1806.00619 [physics.optics]. This report was extremely important as it indicated that equation $P = \frac{2\pi}{k_{SPP}}\sqrt{i^2 + j^2}$ lacks generality. It is safe to say that for an off-normal incident light to have any impact on a particular SPP mode, two conditions, $\mathbf{E}\cdot\mathbf{k}_{\parallel}>0$ and $\mathbf{E}\cdot\mathbf{k}_{SPP}>0$ must be satisfied. All of which indicates that the resilience to incident angle has nothing to do with the aperture being simple circular holes or resonant crosses. Reports such as [23], whose authors (and I don't have any other way of stating this) lacked the basic understanding governing surface waves attributed the resilience in optical response of an array of crosses to the resonance of the aperture, inferring the possibilities to segregate SPP modes from those of the aperture, and in the process have mistaken Wood anomalies with resonant SPPs, as well as few other misunderstanding regarding higher modes…etc. The work presented by authors in [23] is of course just an example of misunderstandings being asserted in the literature as well as being reiterated through other channels, such as dictating to student under supervision, presentations and published derivative works, solely because it was published previously.